\documentstyle[aps,epsfig]{revtex}

\begin{document}
\draft

\title{Parametric instability of linear oscillators with colored
      time-dependent noise}

\author{F.M.Izrailev $^1$, V. Dossetti-Romero $^1$,
A.A.Krokhin$^{1,2}$, and L.Tessieri $^3$}
\address{$^1$ Instituto de F\'{\i}sica, Universidad Aut\'{o}noma de Puebla,
Apdo. Postal J-48, Puebla, Pue. 72570, M\'{e}xico}
\address{$^2$ Center for Nonlinear Science, University of North
Texas, P.O. Box 311427, Denton, Texas 76203-1427, USA}
\address{$^3$ Instituto de F\'isica y Matem\'aticas, 
Universidad Michoacana de San Nicol\'{a}s de Hidalgo, \\
Ciudad Universitaria, 58060 Morelia, Mich. M\'exico}

  \date{\today}
  \maketitle

\begin{abstract}
The goal of this paper is to discuss the link between the quantum
phenomenon of Anderson localization on the one hand, and the
parametric instability of classical linear oscillators with
stochastic frequency on the other. We show that these two problems
are closely related to each other. On the base of analytical and
numerical results we predict under which conditions colored
parametric noise suppresses the instability of linear oscillators.
\end{abstract}

\section{INTRODUCTION}
\label{sect:intro}

Let us consider the one-dimensional (1D) model described by the
Schr\"odinger equation
\begin{equation}
-\frac{\hbar ^{2}}{2m}\psi ^{\prime \prime }(x)+ U(x)\psi
(x)=E\psi (x).  \label{Schr}
\end{equation}
Here the $\psi -$function is a stationary solution for a particle
of energy $E$ moving in a random potential $U(x)$. To simplify the
form of the analytical expressions, in what follows we use energy
units such that $\hbar ^{2}/2m=1$, and we set the zero of the
energy scale so that the mean value of the disordered potential is
zero, $\left\langle U(x)\right\rangle =0$. Here the angular
brackets $\left\langle ...\right\rangle $ denote the average over
the disorder (i.e., over different realizations of $U(x)$). We
restrict our considerations to the case of weak disorder, defined
by the condition $\varepsilon ^{2}=\left\langle
U^{2}(x)\right\rangle \ll 1$.

In the analysis of the model~(\ref{Schr}) one of the main
questions is about global structure of the eigenstates $\psi (x)$
in an infinite configuration space, $ -\infty < x < + \infty$. Of
a particular interest is the problem of whether the eigenstates
are localized or extended for $x\rightarrow \pm \infty $. As was
shown in Ref.~\cite{Ish73}, in 1D models any amount of disorder
(even an infinitesimal one) results in the localization of all
eigenstates (with the exception of a zero-measure set) provided
that the potential $U(x)$ is completely random. This means that
the amplitude of every eigenstate decays exponentially towards
infinity, therefore, far away from the localization center
$x_{0}$, one can write,
\begin{displaymath}
\left| \psi (x)\right| \sim \exp \left( -\left| x-x_{0}\right|
/l\left( E\right) \right) .
\end{displaymath}
Here $l\left( E\right) $ is the so-called localization length that
characterizes, in average, the decrease of the amplitude of the
eigenstate corresponding to the energy $E$. Analytical expression
of $l\left( E\right)$ that is relatively easy to obtain for a weak
disorder, for which it is known that the localization length is
inversely proportional to the square of the disorder strength, $l
\sim 1/\varepsilon^{2}$, see below.

Taking into account that the energy of a free electron is
$E=k^{2}$ , the equation~(\ref{Schr}) can be written in the form
of wave equation,
\begin{equation}
\psi ^{\prime \prime }(x)+k^{2}\psi (x)=U(x)\psi (x)  \label{wave}
\end{equation}
that describes wave propagation in different classical systems.
One example is the propagation of electromagnetic waves in
single-mode waveguides with a rough surface \cite{IM01}. In this
application the potential $U(x)$ is determined by the horizontal
profile $\xi (x)=\varepsilon \varphi (x/R_{c})$, where $R_{c}$ is
the correlation length of the profile, $\varepsilon \ll d$ is the
amplitude of the profile with $d$ being the transverse size of the
waveguide. Note that in this case the parameter $k$ in
Eq.(\ref{wave}) has the meaning of the longitudinal wave number
$k=\sqrt{(\omega/c)^2-(\pi /d)^2}$ where $\omega$ is the frequency
of the wave.

\section{Discrete models}

When the potential is constituted by a succession of delta
scatterers, the model~(\ref{wave}) takes the specific form
\begin{equation}
\psi ^{\prime \prime }(x)+k^{2}\psi (x)=\sum_{n=-\infty
}^{n=\infty }U_{n}\psi (x_{n})\delta (x-x_{n}).  \label{discrete}
\end{equation}%
Here $U_{n}$ is the amplitude of the $n$th delta-scatterer
situated at $x=x_{n}$. In experiments, potentials of this kind are
quite easy to construct; in particular, one can obtain a
realization of delta scatterers by inserting an array of screws
with predetermined lengths and appropriate positions in a
single-mode waveguide~\cite{KIKS00}. Typically, one considers two
limit cases. The first one occurs when all amplitudes $U_{n}$ are
random variables, while the scatterers are periodically spaced,
i.e., $x_{n}=an$ . In this case one can speak of {\it amplitude
disorder}. The second case is represented by the opposite
situation in which the amplitudes of the scatterers are constant,
$U_{n}=U_{0}$ while the positions $x_{n}$ are randomly distributed
around their mean values, i.e., $x_{n}=an+\eta_{n}$ with
$\left\langle \eta_{n} \right\rangle = 0$ and $\left\langle
\eta_{n}^{2}\right\rangle \ll a^{2}$ . Clearly, in the latter case
the mean value of the potential is not zero; however one can
handle this case within the framework of zero-mean potentials by
making use of the special transformation to new variables, see
details in Ref~\cite{IKU01}. This second limit case can be
referred to as {\it positional disorder}.

Due to the delta-like form of a random potential, the
model~(\ref{discrete}) can be considered as a {\it discrete} one.
In fact, its analysis can be reduced to the study of an equivalent
classical two-dimensional map which can be obtained by integrating
Eq.~(\ref{discrete}) between two successive 创kicks创 of the
scattering potential~\cite{IKU01},
\begin{equation}
\begin{array}{cc}
 p_{n+1}=(p_{n}\,+A_{n}\,q_{n})\cos \mu _{n}\,\,-\,\,q_{n} \sin
 \mu _{n}, &
\\
 q_{n+1}=(p_{n}\,+A_{n}\,q_{n})\sin \mu _{n}\,\,+\,\,q_{n} \cos
 \mu_{n} \, .
\end{array}
\label{map}
\end{equation}
Here $q_n$ and $p_n$ are conjugate coordinates and momenta defined
by the identities
\begin{eqnarray*}
q_{n} = \psi_{n}  & {\mbox and } &
p_{n} = (\psi_{n} \cos \mu_{n-1} - \psi_{n-1})/\sin \mu_{n-1}
\end{eqnarray*}
where $\psi_{n}$ is the value of the $\psi-$function at the
position $x=x_n$. The parameter $\mu_n$ is the phase shift of the
$\psi-$function between two scatters,
\begin{equation}
\mu_{n} = k (x_{n+1}-x_{n}) \label{mu_n}
\end{equation}
and the amplitude $A_{n}$ of the $n$th 创kick创 is given by the the
value of the potential at the position $x_n$,
\begin{equation}
A_n= U_n/k \label{A_n}.
\end{equation}
Free rotation in (\ref{map}) between two successive kicks
corresponds to free propagation between scatterers, and each kick
is due to the scattering from a $\delta$ spike of the potential.

In the case of amplitude disorder, the phase shift between two
successive scatterers is the same, $\mu_{n}=\mu=ka$, and the
model~(\ref{discrete}) is known as the Kronig-Penney model. In
this case the two-dimensional map~(\ref{map}) is equivalent to the
following relation between $\psi_{n+1}, \psi_{n-1}$ and
$\psi_{n}$,
\begin{equation}
\psi _{n+1}  + \psi _{n-1} = \left( 2\cos\mu +
 \frac{U_{n}}{k}\sin \mu \right) \psi _{n},
\label{discr}
\end{equation}
One can see that the relation~(\ref{discr}) has the same form as
discrete Schr\"{o}dinger equation for the standard 1D Anderson
tight-binding model,
\begin{equation}
\psi _{n+1}  + \psi _{n-1} = \left(E + \epsilon_n \right)
\psi_{n}, \label{anderson}
\end{equation}
and describes electrons on a discrete lattice with the site
energies $\epsilon_n$. Therefore, many of the results for the
Kronig-Penney model can be obtained by a formal comparison with
the Anderson model, as discussed below.

\section{The Hamiltonian map approach}

One of the tools to find the localization length for discrete
disordered models, either analytically or numerically, is based on
the transfer matrix method. In this approach the localization
length can be expressed as the inverse of the Lyapunov exponent
$\lambda$ which characterizes the growth of the eigenstates
$\psi(x)$ of the stationary Schr\"{o}dinger equation for
increasing $x$. An alternative approach can be obtained by
interpreting the stationary Schr\"{o}dinger equation as the
equation of motion of a classical particle (in this scheme the
space coordinate $x$ of the disordered model is to be seen as the
time coordinate for its dynamical counterpart). In particular, in
the case of discrete disordered models, this approach leads to the
study of classical maps.

It is instructive to illustrate this approach by discussing its
application to the simplest case of the Anderson model
(\ref{anderson}). Comparing Eq.~(\ref{anderson}) with
Eq.~(\ref{map}), one can obtain that there is an exact
correspondence between them by letting $\mu_n=\mu$ and
\begin{equation}
E=2\cos \mu;\,\,\,\,\,\,\, A_n = - \epsilon_n / \sin \mu .
\label{corresp}
\end{equation}
It is clear that for weak disorder the energy spectrum of the
Anderson model~(\ref{anderson}) is close to the unperturbed one
which is defined by the condition $\left| E \right| \leq 2$; this
legitimates the first equality in Eq.~(\ref{corresp}).

To analyze the dynamics of the two-dimensional map~(\ref{map}), it
is convenient to introduce the action-angle variables $(r_n,\theta
_n)$ according to the standard transformation, $q=r\sin \theta
,\,p=r\cos \theta $. As a result, the map gets the following form,
\begin{equation}
\begin{array}{ccl}
\sin \theta_{n+1} & = & D_n^{-1} \left(\sin (\theta_n - \mu) - A_n
\sin \theta_n \sin \mu \right) \\
\cos \theta_{n+1} & = & D_n^{-1} \left(\cos (\theta_n - \mu) + A_n
\sin \theta_n \cos \mu \right)~,
\end{array}
\label{Polar}
\end{equation}
where
\begin{displaymath}
D_n=\frac{r_{n+1}}{r_n}=\sqrt{1+A_n\sin (2\theta
_n)+A_n^2 {\sin}^2\theta _n}~.
\end{displaymath}
Note that the following results for the localization length do not
depend on the sign of $\mu$. It is important that the equation for
the angle $\theta_n$ can be written in the form of the
one-dimensional map,
\begin{equation}
\cot (\theta_{n+1} +\mu) = \cot \theta_n+A_n \label{1d}.
\end{equation}
This fact simplifies the analysis of the distribution of
$\theta_n$. The localization length $l$ is defined as the inverse
Lyapunov exponent, and the latter is determined by the standard
relation \cite{LGP88}
\begin{displaymath}
l^{-1}=\lambda = \lim _{N\to \infty }\langle \frac
1N\sum_{n=0}^{N-1}\ln \left| \frac{q_{n+1}}{q_{n}}\right| \rangle
=\overline{\langle \ln \left| \frac{q_{n+1}}{q_{n}}\right| \rangle}~.
\end{displaymath}
Here the overbar stays for time average and the brackets for the
average over different disorder realizations. The expression for
$l^{-1}$ can be splitted in two terms
\begin{equation}
\label{loc}l^{-1}=\overline{\langle \ln \left(
\frac{r_{n+1}}{r_n}\right) \rangle }+\overline{\langle \ln \left|
\frac{\sin \,\theta _{n+1}}{\sin \,\theta _n}\right| \rangle }~.
\end{equation}
The second term on the r.h.s. is negligible because it is the
average of a bounded quantity. It becomes important only when the
first term is also small, i.e. at the band edge $\left |E \right|
\approx 2$ or $\mu \approx 0$. Thus, apart from this specific
case, the localization length can be evaluated from the
map~(\ref{Polar}) using only the dependence of the radius $r_n$ on
discrete 创time创 $t_n=n$. It is important to note that the ratio
$ r_{n+1}/r_n$ depends only on the angle $\theta _n$ and not on
the radius $r_n$; as a consequence, one can compute the
localization length just by averaging the first term
in~(\ref{loc}) over the invariant measure $\rho (\theta )$
associated with the 1D angular map~(\ref{1d}).

In a direct analytical evaluation of (\ref{loc}) one can write,
\begin{equation}
\label{Measure}l^{-1}=\int P(\epsilon )\int_0^{2\pi }\ln
(D(\epsilon ,\theta ))\,\rho (\theta )\,{d\theta }{d\epsilon }~,
\label{loc2}
\end{equation}
where $P(\epsilon )$ is the density of the distribution of
$\epsilon _n$, and $\rho (\theta )$ is the invariant measure for
the angle variable. We use here the fact that $\rho (\theta )$
does not depend on the specific sequence $ \epsilon _n$, but can
depend on the moments of $P(\epsilon)$, particularly on its second
moment $\sigma ^2$ (see discussion in \cite{IRT98}). As one can
see, in order to evaluate the expression~(\ref{loc2}), first one
has to determine the invariant measure $\rho (\theta )$.

In the case of weak disorder and not close to the band edges we
have $\left| A_n \right| \ll 1$ and one can use the standard
perturbation theory. This allows one to cast Eq.~(\ref{loc2}) in
the form
\begin{equation}
\label{smalllyap}l^{-1}=\frac{1}{2\sin {}^2\mu }\int \epsilon ^2
P(\epsilon )\,d\epsilon \int\limits_0^{2\pi }\rho (\theta )\left(
\frac 14-\frac 12\cos (2\theta )+\frac 14\cos (4\theta )\right)
d{\theta }~.
\end{equation}
This expression is valid for all energies within the band, but
fails at the band edges, where one must include the contribution
of the second term of Eq.~(\ref{loc}) in the computation of the
inverse localization length (see Ref.\cite{IRT98}). One can also
see from Eq.~(\ref{1d}) that in the limit of weak disorder the map
for $\theta_n$ has the approximate form,
\begin{equation}
\label{smallmap}\theta _{n+1}=\theta _n-\mu -A_n\sin {}^2\theta
_n+A_n^2\sin {}^3\theta _n\cos \,\theta
_n\,\,\,\,\,\,\,\,\,\,\,\,\,\,\,\{\mbox{mod}\,\,2\pi \}.
\end{equation}
Therefore, in the first approximation the invariant measure for
$\theta_n$ is flat and this makes possible an explicit evaluation
of the expression~(\ref{smalllyap}). Thus, one easily obtains
\begin{equation}
\label{standard}l^{-1}=\frac{\sigma ^2}{8\sin ^2 \mu}=\frac{\sigma
^2}{8\sqrt{1-\frac{E^2}{4}}}
\label{loc3}
\end{equation}
where $\sigma^2 = \left< \epsilon_n^2 \right >$ is the variance of
the disorder. It is interesting to note that the
expression~(\ref{loc3}) is not correct at the band center, i.e.,
for $E=0$ (see discussion and references in Ref~\cite{IRT98}). The
reason is that for this energy the standard perturbation theory
fails and one must use specific methods to obtain the correct
expression of $l^{-1}$. As was found, the anomaly at the band
center originates from the fact that for $E = 0$ the density $\rho
(\theta )$ is not flat, instead, it has a slight modulation with
$\theta$. This additional $\theta$-dependence of the invariant
measure is due to the special circumstance that the case $E=0$
corresponds to $\mu=\pi/2$ so that the map~(\ref{smallmap})
becomes approximatively periodic of period four. As a consequence,
$\rho(\theta)$ has a weak modulation of period $\pi/2$ and,
therefore, the fourth harmonic $\cos (4\theta)$ in the
expression~(\ref{smalllyap}) also gives a contribution.

Due to the analogy between the Anderson model~(\ref{anderson})
(A-model) and the Kronig-Penney model~(\ref{discrete}) (KP-model)
with $\mu_n=\mu$, one can derive from the result~(\ref{loc3}) the
expression for the localization length of the KP-model
\begin{displaymath}
l^{-1}(E)=\frac{\varepsilon _0^2}{8k^2}\frac{\sin
^2(ka)}{\sin ^2\gamma }.
\end{displaymath}
Here the phase $\gamma $ $(\,0\leq \gamma \leq \pi )$ is given by
the equation,
\begin{displaymath}
2\cos (ka)+\frac{\varepsilon }k\sin (ka)=2\cos \gamma.
\end{displaymath}
This equation is the well-known dispersion relation for the
periodic Kronig-Penney model; the parameter $\gamma$ plays the
role of the Bloch number.

\section{Anderson localization and parametric instability}
\label{sect:correspondence}

It is easy to see that the Schr\"{o}dinger equation~(\ref{discrete}) for
the quantum 1D disordered model can be interpreted as the dynamical
equation of a linear classical oscillator with a parametric perturbation
constituted by a succession of delta-kicks at times $t_n = x_n$. In
particular, the map~(\ref{map}) corresponding to the Kronig-Penney model
($\mu_n=\mu$) can be obtained by integrating the dynamical equations
between two successive kicks for a stochastic oscillator with Hamiltonian
of the form
\begin{equation}
H = \omega \left( \frac{q^2}{2} + \frac{p^2}{2} \right) +
\frac{q^2}{2} \left( \sum_{n=-\infty}^{\infty} A_{n} \delta(t-nT)
\right). \label{kicks}
\end{equation}
Therefore, $q_{n}$ and $p_{n}$ in Eq.~(\ref{map}) stand for the
position and momentum of the oscillator immediately before the
$n$th kick of amplitude $A_n$ occurring at the time $t=nT$.
Correspondingly, the phase shift between two successive kicks is
given by $\mu=\omega T$ where $\omega$ is the unperturbed
frequency of the oscillator and $T$ is the period between the
kicks.

In this description the exponential localization of the
eigenstates of Eq.~(\ref{discrete}) corresponds to a parametric
instability of the stochastic oscillator~(\ref{kicks}). The
instability manifests itself as an exponential divergence of
initially nearby orbits (orbit instability) and, correspondingly,
as an exponential growth of the average energy of the parametric
oscillator (energy instability). The Lyapunov exponent $\lambda$,
which gives the inverse localization length in the solid-state the
divergence of classical trajectories (or, the rate of the energy
growth).

In the previous section we have considered the case of a weak {\it
uncorrelated} disorder which is characterized by its variance
$\sigma^2$ only. In application to classical oscillators this
corresponds to a {\it white noise} perturbation. In the following,
we consider the general case of {\em colored noise} and show that
noise correlations can lead to a quite unexpected phenomenon. To
discuss the effects of correlated noise in parametric oscillators,
we apply the approach of Ref.~\cite{TI01} to the continuous model
described by the Hamiltonian,
\begin{equation}
H = \omega \left( \frac{q^2}{2} + \frac{p^2}{2} \right) +
\frac{q^2}{2} \xi(t) \label{randosc}
\end{equation}
where $\xi(t)$ is a continuous and stationary noise. This model is
slightly different from the one defined by Eq.~(\ref{kicks})
because the noise $\xi(t)$ is a continuous function of time rather
than a succession of $\delta$-kicks. We assume that the noise
$\xi(t)$ has zero average and that its binary correlator is a
known function,
\begin{equation}
\begin{array}{ccc}
\langle \xi(t) \rangle = 0 & \mbox{and} & \langle \xi(t)
\xi(t+\tau) \rangle = \chi (\tau) .
\end{array}
\label{noisprop}
\end{equation}
Here and below, in contrast to previous sections the symbol
$\langle \ldots \rangle$ will refer to the time average, $\langle
f(t) \rangle = \lim_{T_{0} \rightarrow \infty} \frac{1}{T_{0}}
\int_{0}^{T_{0}} f(t) dt $, which is assumed to coincide with the
ensemble average for the process $\xi(t)$.

We define the Lyapunov exponent as follows,
\begin{equation}
\lambda = \lim_{T_{0} \rightarrow \infty} \lim_{\delta \rightarrow
0} \frac{1}{T_{0}} \frac{1}{\delta} \int_{0}^{T_{0}} \ln
\frac{q(t+\delta)}{q(t)} \; dt . \label{lyap}
\end{equation}
As in the previous section, we introduce polar coordinates via the
standard relations $q = r \sin \theta$, $p = r \cos \theta$. This
allows us to represent Eq.~(\ref{lyap}) in the form
\begin{displaymath}
\lambda = \lim_{T_{0} \rightarrow \infty} \frac{1}{T_{0}}
\int_{0}^{T_{0}} \frac{\dot{r}}{r} \; dt .
\end{displaymath}
To proceed further, we consider the equations for the random
oscillator in polar coordinates
\begin{displaymath}
\dot{\theta} = \omega + \xi(t) \sin^{2} \theta ,
\end{displaymath}
\begin{displaymath}
\dot{r} = - \frac{1}{2} r \xi(t) \sin 2 \theta .
\end{displaymath}
Using the last equation, the expression for the Lyapunov exponent
can be finally written in the form
\begin{equation}
\begin{array}{ccl}
\displaystyle \lambda & = & \displaystyle  \lim_{T_{0}
\rightarrow \infty} \frac{1}{2T_{0}} \int_{0}^{T_{0}} \xi(t) \sin
\left( 2 \theta(t) \right) \; dt  \displaystyle = \frac{1}{2}
\langle \xi(t) \sin \left( 2 \theta(t) \right) \rangle .
\end{array}
\label{lyap2}
\end{equation}

Therefore, the problem of computing the Lyapunov
exponent~(\ref{lyap}) is reduced to that of calculating the
noise-angle correlator that appears in Eq.~(\ref{lyap2}). This was
done in Ref.\cite{TI01} by extending the procedure, originally
introduced in Ref.~\cite{IK99} for discrete models, to the
continuum case. As a result, the expression for the Lyapunov
exponent takes the simple but non-trivial form,
\begin{equation}
\lambda = \frac{1}{8} \int_{-\infty}^{+\infty} \langle \xi(t)
\xi(t+\tau) \rangle \cos (2 \omega \tau) \; d \tau \label{lyap3}.
\end{equation}
One can see that the Lyapunov exponent for the stochastic
oscillator~(\ref{randosc}) is proportional to the Fourier
transform $\tilde{\chi}(2\omega)$ of the correlation function at
twice the frequency of the unperturbed oscillator.

A similar result can be obtained for the parametric oscillator~(\ref{kicks})
with discrete noise. In this case the inverse localization length can be
written as~\cite{IK99}
\begin{equation}
\lambda = \frac{\langle A_{n}^{2} \rangle}{8T}\varphi \left(
\omega T \right).
\label{discrlyap}
\end{equation}
Here the function $\varphi \left( \omega T \right)$ is the Fourier
transform,
\begin{displaymath}
\varphi \left( \omega T \right) =  1 + 2 \sum_{k=1}^{+\infty}
\zeta (k) \, \cos \left( 2 \omega T k \right)
\end{displaymath}
of the binary correlator
\begin{equation}
\zeta(k) = \frac{ \langle A_{n+k} A_{n} \rangle}{\langle A_{n}^{2}
\rangle}
\label{bincor}
\end{equation}
of the colored noise. Therefore, the final expression is given by
the product of two factors, namely, the Lyapunov exponent for
the white noise case and the function $\varphi (\omega T)$, which
describes the effect of the noise correlations (the color). In the
case of white noise we have $\varphi (\omega T)=1$.

\section{Suppression of the parametric instability}
\label{sect:mobility}

Expressions~(\ref{lyap3}) and~(\ref{discrlyap}) of the Lyapunov
exponent for stochastic oscillators with weak frequency noise give
a remarkable result: within the limits of the second-order
approximation the rate of parametric instability depends only on
the binary correlator of the noise. In application to solid state
models this fact has suggested a way to construct random
potentials with specific spatial correlations that result in
``windows of transparency'' in the energy spectrum. Indeed, if the
Lyapunov exponent vanishes within some range of the energy (or,
the wave number $k$), then the corresponding eigenstates are
extended in that energy interval. When one considers finite
samples, this means that the transmission coefficient has to be
one in the energy windows where the Lyapunov exponent vanishes.

The possibility of engineering random potentials in order to
obtain Lyapunov exponents with predefined energy dependence can be
deduced from the expression~(\ref{lyap3}) for continuous model, or
from Eq.~(\ref{discrlyap}) for the discrete one. Both expressions
show that if the Lyapunov exponent is known, the two-point
correlator of the corresponding noise can be computed with an
inverse Fourier transform. Since a stochastic process is not
completely determined by its two-point correlator, one can
conclude that there is actually an infinite set of noises which
give rise to the same Lyapunov exponent because they have an
identical binary correlator.

As one can see, in order to have suppression of the parametric
instability in classical oscillators with colored noise, one needs
to have $\lambda (\omega)=0$ in some range of $\omega$. Although
at first sight the construction of a random potential $A_n$ or
$\xi(t)$ with a given binary correlator seems a difficult task, a
rather simple method to solve this problem was presented in
Ref.~\cite{IK99} for discrete models. This method was subsequently
extended to oscillators with continuous noise. Here we describe
how the method works for both classes of oscillators. Let us
consider the continuous model~(\ref{randosc}) first. The starting
point is the correlation function $\chi(\tau)$ that can be easily
obtained by inverting formula~(\ref{lyap3}),
\begin{displaymath}
\chi(\tau) = \frac{8}{\pi} \int_{-\infty}^{\infty} \lambda(\omega)
e^{2i \omega \tau} \; d \omega .
\end{displaymath}
Once the correlation function $\chi(\tau)$ is known, we can obtain
a stochastic process $\xi(t)$ satisfying the
conditions~(\ref{noisprop}) by means of the convolution product,
\begin{equation}
\xi(t) = \left( \beta \ast \eta \right) (t) =
\int_{-\infty}^{+\infty} \beta(s) \eta(s+t) \; ds , \label{conv}
\end{equation}
where the function $\beta(t)$ is related to the Fourier transform
$\tilde{\chi} (\omega)$ of the noise correlation function through
the formula
\begin{displaymath}
\beta(t) = \int_{-\infty}^{+\infty} \sqrt{\tilde{\chi} (\omega)}
e^{i \omega t} \; \frac{d \omega}{2 \pi},
\end{displaymath}
with $\eta(t)$ as any stochastic process such that
\begin{equation}
\begin{array}{ccc}
\langle \eta(t) \rangle = 0 & \mbox{and} & \langle \eta(t)
\eta(t') \rangle = \delta (t-t') .
\end{array}
\label{eta}
\end{equation}
Formula~(\ref{conv}) defines the family of noises corresponding to
a specific form $\lambda (\omega)$ of the frequency-dependent
Lyapunov exponent and gives the solution of the ``inverse
problem'' (i.e., determination of a noise $\xi(t)$ that generates
a pre-defined Lyapunov exponent).

As an example, we consider the Lyapunov exponent
\begin{equation}
\lambda (\omega) = \left\{
                   \begin{array}{cc} 1 & \mbox{if} \;\; \;|\omega|<1/2 \\
                                     0 & \mbox{otherwise}
                   \end{array} \right. ,
\label{edge}
\end{equation}
whose frequency dependence implies that the random oscillator
undergoes a sharp transition for  $|\omega|=1/2$, passing from an
energetically stable condition to an unstable one. Following the
described procedure, it is easy to see that the Lyapunov
exponent~(\ref{edge}) is generated by a noise of the form
\begin{displaymath}
\xi(t) = \frac{\sqrt{8}}{\pi} \int_{-\infty}^{+\infty}
\frac{\sin(s)}{s} \eta(s+t) \; ds ,
\end{displaymath}
with $\eta(t)$ being any random process with the statistical
properties~(\ref{eta}). Working along these lines, it is easy to
see that one can also construct the frequency noises such that the
parametric oscillator is stable for all values of $\omega$ except
those contained in a narrow frequency window.

We now turn our attention to discrete models of the
form~(\ref{kicks}). To show how the expression that is equivalent
to Eq.~(\ref{conv}) can be worked out for this class of systems,
we refer to the case of the Anderson model~(\ref{anderson}) with
correlated disorder. Since this model can be put into one-to-one
correspondence with the kicked oscillator~(\ref{kicks}), it is
perfectly legitimate to analyse each of the two models in terms of
the other; this approach has also the advantage of enhancing the
physical understanding of the problem because it allows one to
interpret the parametric instability of a stochastic oscillator in
terms of localization of electronic states for the Anderson model.

When we transpose the result~(\ref{discrlyap}) to the case of the
Anderson model~(\ref{anderson}), we obtain that the expression for the
localization length has the form,
\begin{equation}
l^{-1}=\frac{\left< \epsilon_{n}^{2}
\right>}{8\sin^{2}\mu}\varphi(\mu), \label{eq:lcorrphi}
\end{equation}
where
\begin{displaymath}
\varphi(\mu)=1\;+\;2\sum_{k=1}^{\infty}\zeta (k)\cos(2\mu k).
\end{displaymath}
Here $\zeta(k)$ is the binary correlator~(\ref{bincor}) which can
be written in terms of the site energies $\epsilon_{n}$ of the
Anderson model as $\zeta (k)= \langle
\epsilon_{n}\epsilon_{n+k}\rangle/
\langle\epsilon_{n}^{2}\rangle$.

If the Lyapunov exponent~(\ref{eq:lcorrphi}) (therefore, the
function $\varphi(\mu)$) is known, the binary
correlators~(\ref{bincor}) can be derived with an inverse Fourier
transform,
\begin{equation}
\zeta (k) =\frac{2}{\pi}\int_{0}^{\pi/2}\varphi(\mu)\;
\cos(2\mu k)\;d\mu. \label{eq:xi_int}
\end{equation}
As for the continuous model, the sequence of site energies
$\epsilon_{n}$ with the correlator of the specific
form~(\ref{eq:xi_int}) can then be constructed with the
convolution product,
\begin{equation}
\epsilon_{n}=\sqrt{\langle \epsilon_{n}^{2}\rangle}
\sum_{k=-\infty}^{\infty}\beta_{k}Z_{n+k},
\label{eq:epsilonsum}
\end{equation}
where
\begin{displaymath}
\beta_{k}=\frac{2}{\pi}\int_{0}^{\pi/2}\sqrt{\varphi(\mu)}\;
\cos(2\mu k)\;d\mu
\end{displaymath}
and $Z_{n}$ are random numbers with the zero mean and unitary variance.
It is easy to check that the correlators of the site
potential~(\ref{eq:epsilonsum}) coincide with the Fourier
coefficients~(\ref{eq:xi_int}).

As an illustration of the method, we construct the random
potential which results in the following function $\varphi(\mu)$
for the Lyapunov exponent of the discrete Anderson model,
\begin{displaymath}
\varphi(\mu)=\left\{\begin{array}{ccl}
                 C_{0}^{2} & \mbox{ if}& \mu_{1} < \mu < \mu_{2},\\
                 0 & \mbox{if} & 0 < \mu < \mu_{1} \mbox{ or }
                 \mu_{2} < \mu < \pi/2.
                       \end{array} \right.
\end{displaymath}
Here, $C_{0}^{2}=\pi/2(\mu_{2}-\mu_{1})$ is the normalization
constant that results from the condition $\zeta_{0}=1$. The
corresponding localization length exhibits two sharp {\em mobility
edges} at the values $E_{1} = 2\cos\mu_{1}$ and $E_{2} =
2\cos\mu_{2}$. Specifically, in the energy window $E_{1} < E <
E_{2}$ the eigenstates are strongly localized, while they are
extended outside of this window. The binary correlators $\zeta
(k)$ for a random potential resulting in such a situation, are
given by
\begin{displaymath}
\zeta(k)=\frac{C_{0}^{2}}{\pi k}
\left[\sin(2k\mu_{2})-\sin(2k\mu_{1})\right] .
\end{displaymath}

As a result, the expression for the inverse localization length reads
\begin{displaymath}
\lambda =
l^{-1}=\frac{\pi\,\sigma^{2}}{16\sin^{2}\mu\left[\mbox{acos}
\left(\frac{E_{2}}{2}\right) - \mbox{acos}\left(\frac{E_{1}}{2}\right)\right]}
\end{displaymath}
where $\sigma^{2}= \left < \epsilon_n^2 \right>$. If the energy
window $\Delta_{E}=E_{1}-E_{2}$ is narrow, one can write,
\begin{displaymath}
\arccos\left(\frac{E_{2}}{2}\right)-\arccos\left(\frac{E_{1}}{2}\right)
\approx \Delta_{E} \; .
\end{displaymath}
One can see that the narrower the window $\Delta_{E}$, the sharper
the transition which occurs at the mobility edges. This effect can
be easily observed numerically, and it may have interesting
applications for parametric oscillators. Indeed, small
localization lengths correspond to large values of the Lyapunov
exponent. Therefore, for values of the frequency of the kicked
oscillator~(\ref{kicks}) which correspond to energy values within
the localization window in the related Anderson model, the
instability of the oscillator is very strong and one can speak of
a kind of  ``parametric stochastic resonance".

\section{Discussion}

In the previous section we have shown how a proper choice of
colored noise (random potentials in the solid state models) can
suppress the parametric instability of a stochastic oscillator in
a prescribed frequency range. One should note, however, that the
theoretical analysis has been focused on the case of weak noise
and that almost all analytical results have been obtained using
perturbative approach. Therefore, the conclusion that the Lyapunov
exponent can vanish within some frequency region is valid only
within the framework of second-order perturbation theory. Going
beyond the second-order approximation, it is possible to estimate
the correction to the present results for the inverse localization
lenght and to show that the correction is represented by a term of
order $O(\sigma^4)$ (with $\sigma^2= \langle \epsilon_{n}^{2}
\rangle$). It is not clear whether one can make this fourth-order
correction vanish with an appropriate choice of the statistical
properties of the noise~\cite{T02}; from a practical point of
view, however, in the case of weak noise there is a well-defined
separation between the time scale $t \sim 1/\sigma^{2}$ over which
the suppression of instability holds, and the much longer time
scale $t \sim 1/\sigma^{4}$ over which the effects of fourth-order
corrections become relevant. When the second-order results for the
inverse localization length are applied to wave-guides or solid
state models, fourt-order effects can also be generally avoided by
a proper choice of size of an experimental device~\cite{KIKS00}.

The transfer matrix method in the Hamiltonian form described above
is also very useful for finite times. In application to solid
state models this question refers to transport properties through
finite samples of size $L$. As is known~\cite{LGP88,KTI97}, all
transport properties can be directly related to the classical
trajectories of the Hamiltonian map~(\ref{map}). Specifically, by
studying general properties of these trajectories, one can find
statistical properties of the transmission coefficient or the
resistance. The transmission coefficient through a $L$-site sample
can be expressed in terms of dynamical variables of the classical
map~(\ref{map}) as
\begin{displaymath}
T_{L} = \frac{4}{2+r_{1}^{2}+r_{2}^{2}}
\end{displaymath}
where $r_{1}$ and $r_{2}$ represent the radii at the $L$th step of
the map trajectories starting from the phase-space points
$P_{1} = (x_{0}=1,p_{0}=0)$ and $P_{2} = (x_{0}=0,p_{0}=1)$, respectively.
As for the resistance $R_{L}$, it is defined as the inverse of the
transmission coefficient $R_{L}=T_{L}^{-1}$.
The key feature of these formulae is that they express the transport
properties of a disordered sample in terms of the radii of map
trajectories in the phase space.
On the other hand, the square radius $r^{2}$ of a map trajectory is a
quadratic function of the coordinate and momentum of the corresponding
kicked oscillator, $r^{2} = p^{2} + q^{2}$, and is therefore proportional
to the energy of the latter.
This fact makes possible to relate transport properties of quantum models
with the time dependence of the energy of classical parametric oscillators.

It is possible to obtain quite easily the moments of the energy
$r^{2}$ of the parametric oscillator described by the
Hamiltonian~(\ref{kicks}), see details in Refs.~\cite{TI01,DIK04}.
In particular, one can obtain that in the asymptotic limit (i.e.,
for times $t \gg \lambda^{-1}$) the mean value of the energy grows
exponentially as
\begin{displaymath}
\langle r^{2}(t) \rangle = r^{2}(0) \exp(4\lambda t)
\end{displaymath}
where $\lambda$ is the Lyapunov exponent~(\ref{discrlyap}). This
formula shows that the exponential rate of the energy growth for
the parametric oscillator is four times the Lyapunov exponent,
i.e., the rate of exponential separation of nearby
orbits~\cite{TI01}.

Another important question concerns the fluctuations of $r^{2}(t)$
for fixed times $t$ depending on different realizations of the
noise. Using the results of Ref.~\cite{TI01,DIK04}, one should
distinguish between two different situations. The first one
corresponds to small times when the value of $r(t)$ is close to
the initial value $r(0)$. In solid state models this case is known
as the ballistic transport for which the localization length
$l=\lambda^{-1}$ is much larger than the size $L=t$ of the sample,
$\lambda t \ll 1$. Another limit case corresponds to large times,
$\lambda t \gg 1$, or to the strongly localized regime in quantum
models. One of the most interesting effects is that in this case
the fluctuations of the energy of the classical oscillator
(resistance in quantum models) are huge and the quantity $R=r^{2}$
is not self-averaging. To deal with a well-behaved (that is,
self-averaging) statistical property, one has to consider the
logarithm of the oscillator energy, which has a Gaussian
distribution for large times. It turns out that the energy $r^{2}$
has log-normal distribution:
\begin{displaymath}
P(r^{2},t) = \frac{1}{\sqrt{8\pi \lambda t}} \exp
\left[ - \frac{\left( \ln r^{2} -2 \lambda t \right)^{2}}{8\lambda t} \right].
\end{displaymath}
This distribution implies that the energy of parametric
oscillator, or the resistance $R=r^{2}$ of disordered samples,
satisfy the relations
\begin{eqnarray*}
\langle \ln R \rangle = 2 \Lambda; & &
\langle \ln^{2} R \rangle = 4 \Lambda + 4 \Lambda^{2} \; ,
\end{eqnarray*}
\begin{displaymath}
\mbox{Var} \left( \ln R \right) =
\langle \ln^{2} R \rangle -
\langle \ln R \rangle^{2} = 2 \langle \ln R \rangle .
\end{displaymath}
where $\Lambda =\lambda t$.

In conclusion, we have discussed the analogy between properties of
quantum 1D models with random potentials and classical linear
oscillators governed by parametric noise. We have shown that many
results known for quantum models can be mapped unto corresponding
properties of classical oscillators. One of the important
questions is about the time-dependence of the energy of stochastic
oscillators with frequency perturbed by a white noise. Another,
even more exciting problem, is the behavior of the oscillators
when the frequency noise has long-range correlations. It was shown
that in the case of weak noise all statistical properties of the
classical trajectories depend on the binary correlator of the
noise only. This fact opens the door to the construction of
colored noises with specific long-range correlations which result
in a sharp change in the dynamical behavior of the parametric
oscillator at some threshold value of the unperturbed frequency.
Specifically, the characteristic instability of parametric
oscillators can be suppressed in a certain frequency range (with a
brisk transition), thanks to long-range temporal correlations of
the noise. These results may find different applications in the
field of classical systems with colored noise.
\acknowledgments     

The authors are very thankful to N.M.Makarov for fruitful
discussions and valuable comments.


\begin{thebibliography}{1}

\bibitem{Ish73} K. Ishii, {\it Suppl. Progr. Theor. Phys.} {\bf 53},
p.~77, 1973.

\bibitem{IM01} F.~M. Izrailev and N.~M. Makarov,
{\em Optics Lett.} {\bf 26}, p.~1604, 2001.

\bibitem{KIKS00} U. Kuhl, F.~M. Izrailev, A.~A. Krokhin,
and H.-J. St\"ockmann, {\em Appl. Phys. Lett.} {\bf 77}, p.~633,
2000; A.~A. Krokhin, F.~M. Izrailev, U. Kuhl, H.-J. St\"ockmann,
and S. Ulloa, {\em Physica E} {\bf 13}, p.~695, 2002.

\bibitem{IKU01} F.~M. Izrailev, A.~A. Krokhin, and S.~E. Ulloa,
{\em Phys. Rev. E.} {\bf 63}, p.~041102, 2001.

\bibitem{LGP88} I.~M. Lifshitz, S. Gredeskul, and L. Pastur,
{\em Introduction to the Theory of Disordered Systems}, Wiley, New
York, 1988.

\bibitem{IRT98} F.~M. Izrailev, S. Ruffo and L. Tessieri, {\em J. Phys. A: Math.
Gen.} {\bf 31}, p.~5263, 1998.

\bibitem{TI01} L. Tessieri and F.~M. Izrailev, {\em Phys. Rev. E} {\bf 64},
p.~66120, 2001.

\bibitem{IK99} F.~M. Izrailev and A.~A. Krokhin, {\em Phys. Rev. Lett.}
{\bf 82}, p.~4062, 1999.

\bibitem{T02} L. Tessieri, {\em J. Phys. A: Math.
Gen.} {\bf 35}, p.~9585, 2002.

\bibitem{KTI97} T. Kottos, G.~P. Tsironis and F.~M. Izrailev, {\em J. Phys.: Condens.
Matter} {\bf 9}, p.~1777 , 1997.

\bibitem{DIK04} V. Dossetti, F.~M. Izrailev, and A.~A. Krokhin,
{\em Phys. Lett. A} {\bf 320}, p.~276, 2004.

\end{thebibliography}
\end{document}